# Flexocoupling impact on the size effects of piezo- response and conductance in mixed-type ferroelectrics-semiconductors under applied pressure

*Anna N. Morozovska*[1,1], *Eugene A. Eliseev*[2], *Yuri A. Genenko*[3], *Ivan S. Vorotiahin*[1, 3], *Maxim V. Silibin*[4], *Ye Cao*[5], *Yunseok Kim*[6], *Maya D. Glinchuk*[2], *and Sergei V. Kalinin*[52]

[1] *Institute of Physics, National Academy of Sciences of Ukraine,*
*46, pr. Nauky, 03028 Kyiv, Ukraine*

[2] *Institute for Problems of Materials Science, National Academy of Sciences of Ukraine,*
*Krjijanovskogo 3, 03142 Kyiv, Ukraine*

[3]*Institut für Materialwissenschaft, Technische Universität Darmstadt, Jovanka-Bontschits-Str. 2,*
*64287 Darmstadt, Germany*

[4]*National Research University of Electronic Technology "MIET", Bld. 1, Shokin Square, 124498*
*Moscow, Russia*

[5] *The Center for Nanophase Materials Sciences, Oak Ridge National Laboratory,*
*Oak Ridge, TN 37831*

[6] *School of Advanced Materials Science and Engineering, Sungkyunkwan University (SKKU),*
*Suwon 16419, Republic of Korea*

**Abstract**

Flexocoupling impact on the size effects of the spontaneous polarization, effective piezo-response, elastic strain and compliance, carrier concentration and piezo-conductance have been calculated in thin films of ferroelectric semiconductors with mixed-type conductivity under applied pressure. Analysis of the self-consistent calculation results revealed that the thickness dependences of aforementioned physical quantities, calculated at zero and nonzero flexoelectric couplings, are very similar under zero applied pressure, but become strongly different under the application of external pressure $p_{ext}$. At that the differences become noticeably stronger for the film surface under compression than under tension. The impact of the Vegard mechanism on the size effects is weaker in comparison with flexocoupling except for the thickness dependence of the piezo-conductance.

Without flexoelectric coupling the studied physical quantities manifest conventional peculiarities that are characteristic of the size-induced phase transitions. Namely, when the film thickness $h$ approaches the critical thickness $h_{cr}$ the transition to paraelectric phase occurs. The combined effect of flexoelectric coupling and external pressure induces polarizations at the film surfaces, which cause the electric built-in field that destroys the thickness-induced phase transition to paraelectric phase at $h = h_{cr}$ and induces the electret-like state with irreversible spontaneous polarization at $h < h_{cr}$. Rather unexpectedly the built-in field leads to noticeable increase of the average strain and elastic compliance under the film thickness decrease below $h_{cr}$ that scales as $1/h$ at small thicknesses $h$. Also the built-in field induces strong and non-trivial thickness dependence of free electron concentration and effective piezo-conductance for all film thicknesses $h$ including the range of small thickness $h \leq h_{cr}$. The changes of the electron concentration by several orders of magnitude under positive or negative pressures can lead to the occurrence of high- or low-conductivity states, i.e. the nonvolatile piezo-resistive switching, in which the swing can be controlled by the film thickness and flexoelectric coupling.

---

[1] corresponding author 1, e-mail: <u>anna.n.morozovska@gmail.com</u>
[2] corresponding author 2, e-mail: <u>sergei2@ornl.gov</u>

Obtained theoretical results can be of fundamental interest for ferroics physics, thin films, semiconductor physics, interferometry and scanning probe microscopy. Predicted non-trivial behavior of the elastic properties and piezo-conductance are waiting for experimental verification.

## I. Introduction

Investigation of electromechanical, electrochemical and electrophysical properties of nanosized ferroelectrics-semiconductors with mixed type ionic-electronic conductivity (**FeMIECs**) is of significant interest for both fundamental science and numerous applications of these multifunctional materials [1, 2]. Although FeMIECs in the form of thin films and nanocomposites are among the most promising MIECs materials for the next generation of nonvolatile, resistive and memristive memories, logic devices, ultrasensitive sensors, miniature actuators and positioners [3, 4, 5], the physical principles of the complex interplay between the ferroelectric polarization, elastic strains, ionic and electronic state at the nanoscale are not clear so far. Naturally the lack of physical understanding precludes the successful implementation of FeMIECs in the aforementioned applications.

Among various unsolved fundamental problems in the field of MIEC physics, we would like to focus on the importance of the theoretical description of the size effect on the local electromechanical, electrochemical and electrophysical properties of nanosized FeMIECs, which is virtually absent to date. The problem appeared since numerous studies of MIECs and FeMIECs local electromechanical responses (**piezo-response**) by Scanning Probe Microscopy (SPM) and electromechanical conductance (**piezo-conductance**) by Electrochemical Strain Microscopy (ESM) [6], Piezoresponse Force Microscopy (PFM) [7] and Current Atomic Force Microscopy (C-AFM) [8] revealed that their electro-conductance is strongly coupled with polar and elastic state via electric field, ion and electron dynamics and mechanical pressure. Moreover, both SPM and interferometric measurements with high sub-nm resolution indicate the important role of the local gradients of polarization, strain and space charge density in the formation of aforementioned local response [6-8]. The local gradient of polarization induces inhomogeneous elastic strain, and vice versa, the gradient of elastic strain induces polarization due to the flexoelectric coupling (**flexocoupling**) [9, 10]. The gradients inevitably cause the space charge redistribution in MIECs and FeMIECs via several mechanisms [11, 12], such as electromigration and diffusion [13, 14], chemical strains and stresses [15, 16, 5], and deformation potential [17, 18].

The flexoelectricity impact is of great importance in nanoscale objects [19, 20], for which the strong strain gradients are inevitably present near the surfaces, in thin films [21, 22, 23], nanoparticles [24] and fine-grained ceramics [25, 26]. Therefore the role of flexocoupling in the formation of piezo-response and piezo-conductance can essentially increase due to the intrinsic size effects, which become pronounced when the thickness of investigated FeMIEC film becomes less than 50 nm, over the circumstances discussed below.

As a rule, intrinsic size effect in thin ferroelectric films manifests itself in the ferroelectric phase disappearance when the film thickness becomes smaller than the critical thickness [27]. The critical thickness value primary depends on the polarization direction, correlation length, surface energy contribution, electrical and mechanical conditions at the film surfaces [28, 29, 30, 31]. The surface energy determines the value of the so-called extrapolation lengths [28]. Depolarization field is originated from nonzero divergence of polarization vector, as well as from the incomplete screening of the polarization bound charges by the electrodes [28, 31]. Elastic strains are caused by e.g. film and substrate lattice mismatch [29, 30]. All these factors, which are closely related to the surface influence, often lead to the appearance of a developed polarization gradient from the film surface towards its center. The polarization gradient induces inhomogeneous elastic strain due to the flexoelectric coupling.

Surface piezoelectric effect coupled with misfit strain leads to the appearance of built-in electric field that in turn destroys the size-induced phase transition into a paraelectric phase at the critical thickness and induces the electret-like state with irreversible polarization at film thickness less than the critical one [29, 30]. Our idea is that applied pressure coupled with the flexoelectric effect can act similarly on misfit strain, wherein the pressure control is much easier experimentally. Moreover, the pressure application to MIECs by SPM indentation can strongly affect its conductivity state (up to a metal−insulator transition mediated by ionic dynamics and ferroic phase transitions) and thus allows the study of material physics and enables novel data storage technologies with mechanical writing and current-based readout [8].

Aforementioned facts motivated the work devoted to the theoretical modeling of the flexocoupling impact on the size effects of the spontaneous polarization, effective piezo-response, elastic strain and compliance, carrier concentration and piezo-conductance in thin films of FeMIECs under applied pressure.

## II. Problem statement and basic equations

Generalized expression for the Landau-Ginzburg-Devonshire (LGD)-type Gibbs potential of the spatially confined ferroelectric mixed-type semiconductors, that is the sum of the bulk ($G_V$) and surface ($G_S$) parts, has the following form [12, 32]:

$$G_V = \int_V d^3r \begin{pmatrix} \dfrac{a_{ik}}{2}P_iP_k + \dfrac{b_{ijkl}}{4}P_iP_jP_kP_l + \dfrac{g_{ijkl}}{2}\left(\dfrac{\partial P_i}{\partial x_j}\dfrac{\partial P_k}{\partial x_l}\right) - P_iE_i - Q_{ijkl}\sigma_{ij}P_kP_l - \dfrac{s_{ijkl}}{2}\sigma_{ij}\sigma_{kl} \\ -F_{ijkl}\sigma_{ij}\dfrac{\partial P_l}{\partial x_k} - \left(\Sigma^e_{ij}\delta n + W^d_{ij}\delta N^+_d\right)\sigma_{ij} + e\varphi(Z_d N^+_d - n) \\ -N^+_d E_d - T S_d[N^+_d] + nE_C - T S_{el}[n] + \dfrac{3k_B T}{2} N_C F_{3/2}\left(\dfrac{E_g + e\varphi}{k_B T}\right) \end{pmatrix} \quad (1a)$$

$$G_S = \sum_m \int_{Sm} \left(\dfrac{A^m_{jk}}{2}P_jP_k + d^{Sm}_{jkl}u^{Sm}_{jk}P_l\right)d^2r \quad (1b)$$

Hereinafter summation is performed over all repeating indexes; $P_i$ is a ferroelectric polarization, $E_i = -\partial\varphi/\partial x_i$ is a quasi-static electric field, $\varphi$ is the electric potential. The coefficients of LGD potential expansion on the polarization powers, $a_{ik} = \alpha^T_{ik}(T - T_c)$ and $b_{ijkl}$ are also called linear and nonlinear dielectric stiffness coefficients. $T$ is the absolute temperature, $T_c$ is the Curie temperature. Elastic stress tensor is $\sigma_{ij}$, $Q_{ijkl}$ is electrostriction tensor, $F_{ijkl}$ is the flexoelectric effect tensor [33]. Hereinafter we consider materials with inversion center in the parent phase (e.g. with cubic parent phase). $g_{ijkl}$ is gradient coefficient tensor, $s_{ijkl}$ is elastic stiffness. Variations of the electron density, and ionized donor concentration are $\delta n(\mathbf{r}) = n(\mathbf{r}) - n_0$ and $\delta N^+_d(\mathbf{r}) = N^+_d(\mathbf{r}) - N^+_{d0}$. Constant values of $n_0$ and $N^+_{d0}$ correspond to stress-free reference state at zero electric field. $e$ is the electron charge, $Z_d$ is the donor ionization degree. Deformation potential tensor is denoted by $\Sigma^e_{ij}$ and Vegard expansion (another name is elastic dipole) tensor is $W^d_{ij}$ [15, 16]. The Vegard tensor $W^d_{ij}$ for donors will be regarded diagonal hereinafter. Only ionized donors (e.g. light impurity ions or oxygen vacancies) are regarded mobile [34]. Mobile acceptors can be considered in a similar way. $E_d$ is the donor level, $E_C$ is the bottom of the conduction band.

Expression for the entropy of ionized donors, considered in the approximation of an infinitely thin single donor level, has the form $S_d[N^+_d] = -k_B\left(N^0_d\left(\dfrac{N^+_d}{N^0_d}\ln\left(\dfrac{N^+_d}{N^0_d}\right) + \left(1 - \dfrac{N^+_d}{N^0_d}\right)\ln\left(1 - \dfrac{N^+_d}{N^0_d}\right)\right)\right)$. The entropy density of electron Fermi gas, considered in the parabolic or effective mass approximation, is $S_{el}[n] = -k_B N_C \int_0^{n/N_C} d\tilde{n} F^{-1}_{1/2}(\tilde{n})$, where $F^{-1}_{1/2}(\xi)$ is the inverse function to the Fermi ½-integral $F_{1/2}(\xi) = \dfrac{2}{\sqrt{\pi}}\int_0^\infty \dfrac{\sqrt{\zeta}d\zeta}{1 + \exp(\zeta - \xi)}$ (see **Appendix A** of **Suppl. Mat.**). $N_C = \left(m_n k_B T/(2\pi\hbar^2)\right)^{3/2}$ is the effective density of states in the conduction band, electron effective mass is $m_n$ [35]. The partial derivative $\partial S_{el}/\partial n = -k_B F_{1/2}^{-1}(n/N_C)$. Approximations for direct and inverse Fermi integrals are $F_{1/2}(\varepsilon) \approx \left(\exp(-\varepsilon) + (3\sqrt{\pi}/4)(4+\varepsilon^2)^{-3/4}\right)^{-1}$ and $F_{1/2}^{-1}(\tilde{n}) \approx (3\sqrt{\pi}\tilde{n}/4)^{2/3} + \ln(\tilde{n}/(1+\tilde{n}))$ correspondingly [36]. The last term, in Eq.(1a), is the electron kinetic energy [32], $F_{3/2}(\xi) = \dfrac{2}{\sqrt{\pi}} \int_0^\infty \dfrac{\zeta\sqrt{\zeta}d\zeta}{1+\exp(\zeta-\xi)}$ is the Fermi 3/2-integral.

$A_{ij}^{Sm}$ is the surface dielectric stiffness at the surface $S_m$, $d_{ijk}^S$ is the surface piezoelectric tensor, $u_{ij}^S$ is the surface strain field, originated from e.g. film and substrate lattice mismatch [29]. The surface piezoeffect could be essential at distances of order 1-5 lattice constants from the film surface [37], although for strong enough film-substrate lattice mismatch it can be the source of thin film self-polarization (see [29, 30] and refs therein). In what follows we will not consider the latter case.

For ferroelectrics with cubic parent phase the term $Q_{ijkl}P_k P_l$ automatically includes piezoelectric contribution, because the polarization change under electric field $E_m$ can be approximated as [38]

$$P_k^t = P_k(E_m) + \varepsilon_0(\varepsilon_{km}^b - \delta_{km})E_m \approx P_k^S + \varepsilon_0(\varepsilon_{km}^f - \delta_{km})E_m. \qquad (2a)$$

Here $P_k^S$ is a spontaneous polarization component, $\varepsilon_0$ is the dielectric permittivity of vacuum, $\delta_{km}$ is a Kroneker symbol and $\varepsilon_{ij}^f$ is the relative dielectric permittivity of ferroelectric that includes a soft-mode related electric field-dependent contribution $\varepsilon_{ij}^{sm}$ and an electric field-independent lattice background contribution $\varepsilon_{ij}^b$ [21]. Consequently, an apparent piezoelectric coefficient becomes [39]

$$d_{ijk} = 2\varepsilon_0(\varepsilon_{km}^f - \delta_{km})Q_{ijml}P_l^S. \qquad (2b)$$

Further we consider a squashed tip or a thin disk electrode placed in an electric contact with a ferroelectric mixed-type semiconductor film clamped to a rigid bottom electrode. One-component polarization $P_3$ is normal to the film surface that corresponds to a tetragonal ferroelectric phase in a c-film. Problem geometry is shown in **Figure 1.** One-dimensional approximation of capacitor geometry is applicable for the problem solution, if the radius of the top disk electrode is much larger than the film thickness.

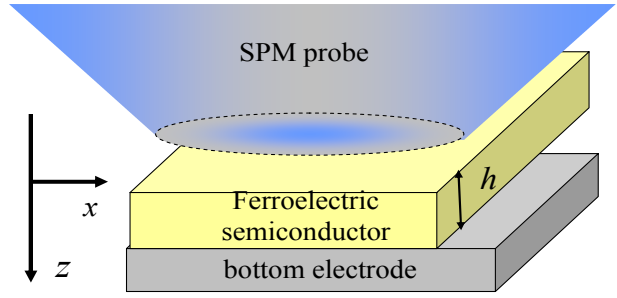

**Figure 1.** Geometry of the considered problem. We consider the situation when either the radius of the SPM tip is much larger than the film thickness, or the ambient screening charges play the role of a top electrode.

For the semiconductor film with mixed ionic-electronic conductivity the electric potential φ can be found self-consistently from the Poisson equation

$$\varepsilon_0 \varepsilon_{33}^b \frac{\partial^2 \varphi}{\partial x_3^2} = \frac{\partial P_3}{\partial x_3} - e\left(Z_d N_d^+(\varphi) - n(\varphi)\right) \qquad (3)$$

with boundary conditions corresponding to the fixed potentials at the electrodes, $\varphi(0) = V$, $\varphi(h) = 0$, including the short-circuited case, $\varphi(0) = \varphi(h) = 0$. In Equation (3) we used the relation (2a) between the total and ferroelectric polarization contributions.

When the system is in thermodynamic equilibrium, currents are absent and electrochemical potentials are equal to the Fermi level. In the considered case donor concentration is $N_d^+ = N_d^0\left(1 - f\left((E_d + W_{ij}^d \sigma_{ij} - eZ_d\varphi + E_F)/k_B T\right)\right)$, where the Fermi-Dirac distribution function is introduced as $f(x) = (1 + \exp(x))^{-1}$ and $E_F$ is the Fermi energy level in equilibrium. Electron density is $n = N_C F_{1/2}\left((e\varphi + \Sigma_{ij}^e \sigma_{ij} + E_F - E_C)/k_B T\right)$, where $F_{1/2}(\xi) = \frac{2}{\sqrt{\pi}} \int_0^\infty \frac{\sqrt{\zeta}\, d\zeta}{1 + \exp(\zeta - \xi)}$ is the Fermi ½-integral.

Inhomogeneous spatial distribution of the ferroelectric polarization component(s) should be determined self-consistently from the LGD-type Euler-Lagrange equations, $\frac{\delta G}{\delta P_i} - \frac{\partial}{\partial x_k}\left(\frac{\delta G}{\delta(\partial P_i/\partial x_k)}\right) = 0$, with boundary conditions at surfaces $S_1$ at $x_3=0$ and $S_2$ at $x_3=h$, $\left.\left(A_{33}^{S1} P_3 - g_{33}\frac{\partial P_3}{\partial x_3} + F_{kl33}\sigma_{kl}\right)\right|_{x_3=0} = 0$ and $\left.\left(A_{33}^{S2} P_3 + g_{33}\frac{\partial P_3}{\partial x_3} - F_{kl33}\sigma_{kl}\right)\right|_{x_3=h} = 0$, which follow from the minimization of the Gibbs potential (1). The conditions are of the third kind due to the flexoelectric effect contribution. The evident form of Euler-Lagrange equations and boundary conditions are listed in the **Appendix B** of **Suppl. Mat.** Remarkably, that the product $F_{kl33}\sigma_{kl}/A_{33}^{Si}$ act as a surface polarization. Note, that the coefficients $A_{33}^{S1}$ and $A_{33}^{S2}$ conditioned by the interface chemistry can be very different for the probed surface $x_3 = 0$, where an active chemical environment can exist, and for the electroded surface $x_3 = h$, where the perfect electric contact is present.

Equation of state for elastic fields, $\delta G/\delta \sigma_{ij} = -u_{ij}$, obtained from the variation of the functional (1), shows that there are four basic contributions to the elastic strain of the spatially-confined ferroelectric materials with mobile charge species, namely purely elastic, flexoelectric, Vegard and electrostriction contributions. Hence the local strain is

$$u_{ij} = s_{ijkl}\sigma_{kl} + F_{ijkl}\frac{\partial P_k}{\partial x_l} + W_{ij}^d\left(N_d^+ - N_{d0}^+\right) + \Sigma_{ij}^e\left(n - n_0\right) + Q_{ijkl}P_k P_l , \qquad (4)$$

The piezoelectric contribution is automatically included in the relation (4) as linearized electrostriction in the ferroelectric phase accordingly to Eq.(2a), and the apparent piezoelectric coefficient $d_{ijk}^{eff}$ can be introduced accordingly to Eq.(2b).

Generalized Hook's relations (4) should be supplemented by the mechanical equilibrium equations $\partial \sigma_{ij}/\partial x_j = 0$ in the bulk and equilibrium conditions $\sigma_{ij}n_j\big|_{S_f} = -p_i^{ext}$ at the free surfaces $S_f$ of the system, $n_j$ is the component of the outer normal $\mathbf{n}=(0,0,-1)$ to the surface $S_f$ [40]. Here we suppose that external pressure $p_i^{ext}$ can be applied to the system. Elastic displacement is zero at the clamped surfaces $S_c$, $u_i(S_c) = 0$. The evident expressions for elastic strains and stresses are listed in the **Appendix B** of **Suppl. Mat.**

The film surface displacement is $u_3 = \int_0^h u_{33}dx_3$ for the considered geometry of the problem. The strain $u_{33}$ is listed in **Appendix B of Suppl. Mat.** The average strain $\langle u_{33}\rangle = \dfrac{u_3}{h}$ has the following form:

$$\langle u_{33}\rangle = -s_{33}^{eff} p_{ext} + W_{33}^{eff}\langle \delta N_d^+\rangle + \Sigma_{33}^{eff}\langle \delta n\rangle + \frac{F_{33}^{eff}}{h}\left(P_3(h) - P_3(0)\right) + Q_{33}^{eff}\langle P_3^2\rangle. \qquad (5)$$

Here we introduced the apparent coefficients $s_{33}^{eff} = s_{33} - \dfrac{2s_{13}^2}{s_{11} + s_{13}}$, $W_{33}^{eff} = W_{33}^d - \dfrac{2s_{13}W_{11}^d}{s_{11} + s_{13}}$, $\Sigma_{33}^{eff} = \Sigma_{33}^e - \dfrac{2s_{13}\Sigma_{11}^e}{s_{11} + s_{13}}$, $F_{33}^{eff} = F_{33} - \dfrac{2s_{13}F_{13}}{s_{11} + s_{13}}$ and $Q_{33}^{eff} = Q_{33} - \dfrac{2s_{13}Q_{13}}{s_{11} + s_{13}}$. Voigt notations are introduced for the electrostriction $Q_{ij}$, gradient coefficient $g_{ij}$, flexoelectric $F_{ij}$ and elastic compliance $s_{ij}$ tensors, while full matrix notations are retained for all other tensors. The tensor components with subscripts 12, 13 and 23 are equal for materials with cubic parent phase. Corresponding effective elastic compliance can be calculated from the formulae:

$$S_{33}^{eff} = -\frac{1}{h}\frac{du_3}{dp_{ext}}. \qquad (6)$$

By definition, effective piezo-response is given by expression $R_3^{eff} = \partial u_3/\partial V$. Since $P_i = -\delta G/\delta E_i$ and $u_{ij} = -\delta G/\delta \sigma_{ij}$, in accordance with Maxwell relations we obtained that

$$R_3^{eff} = \frac{\partial u_3}{\partial V} = \frac{\partial P_3}{\partial p_{ext}} \equiv \frac{\partial^2 G}{\partial V \partial p_{ext}}. \qquad (7)$$

Approximate analytical expressions of effective piezo-response are listed in the **Appendix C** of **Suppl. Mat.**

In order to study the dependence of the film electro-conductance $\Omega$ on applied pressure $p_{ext}$, i.e. the effective piezo-conductance $\Omega_p \equiv d\Omega/dp_{ext}$, one should solve the dynamic problem and calculate a derivative of the electric current with respect to the applied voltage and study this value in dependence on $p_{ext}$, $\Omega_p \equiv \frac{1}{E_{ext}} \frac{\partial J}{\partial p_{ext}}$, here $E_{ext} = V/h$ (assuming linear approximation on $V$). The donor current is $J_d = -eZ_d \eta_d N_d^+ (\partial \zeta_d/\partial x_3)$, where $\eta_d$ is the donor mobility coefficient, $\zeta_d$ is electrochemical potential for donor, $\zeta_d = -(\delta G/\delta N_d^+) \equiv E_d + W_{ij}^d \sigma_{ij} - eZ_d \varphi - k_B T \ln(N_d^+/(N_d^0 - N_d^+))$. The electronic current is $J_e = e\eta_e n(\partial \zeta_e/\partial x_3)$, where $\eta_e$ is the electron mobility coefficient, $\zeta_e$ is electrochemical potential for electron, $\zeta_e = +(\delta G/\delta n) \equiv E_C - \Sigma_{ij}^e \sigma_{ij} + k_B T F_{1/2}^{-1}(n/N_C) - e\varphi$.

Kinetic equations for electrons and donors, $\frac{\partial n}{\partial t} - \frac{1}{e}\frac{\partial J_3^e}{\partial x_3} = 0$ and $\frac{\partial N_d^+}{\partial t} + \frac{1}{eZ_d}\frac{\partial J_d}{\partial x_3} = 0$, are supplemented by ion-blocking boundary conditions $J_d|_{x_3=0,h} = 0$; and fixed electron densities at the electrodes $n(0) = n_0$ and $n(h) = n_1$. For the case of ion-blocking electrodes only electronic current contributes into the conductance $\Omega$. Hence the piezo-conductance can be estimated as (see **Appendix D** of **Suppl. Mat.**):

$$\Omega_p \equiv \frac{1}{E_{ext}}\frac{dJ}{dp_{ext}} \approx e^2 \frac{\eta_e}{h} \frac{d}{dp_{ext}} \int_0^h n\, dx_3 \sim \frac{d\langle n\rangle}{dp_{ext}}, \qquad (8)$$

Below we compare approximate analytical expressions derived in the **Suppl. Mat** with self-consistent numerical modelling with and without flexoelectric coupling and pressure application.

### III. Results of self-consistent calculations and discussion

Size effects of the spontaneous polarization, effective piezo-response, average elastic strain and compliance, electron concentration and piezo-conductance have been calculated in a self-consistent way for PbZr$_{0.5}$Ti$_{0.5}$O$_3$ (PZT) at room temperature (RT). Parameters used are listed in the **Table I.**

**Table I.** Material parameters collected and estimated from the Refs [16, 41, 42]

| coefficient | PbZr$_{0.5}$Ti$_{0.5}$O$_3$ |
|---|---|
| $\varepsilon_b$ | 10 |
| $\alpha^T$ (×10$^5$C$^{-2}$·Jm/K) | 2.66 |
| $T_C$ (K) | 666 |
| $b_{ij}$ (×10$^8$C$^{-4}$·m$^5$J) | $b_{33}$= 3.98 |
| $Q_{ij}$ (C$^{-2}$·m$^4$) | $Q_{33}$=$Q_{11}$=0.0812, $Q_{13}$= −0.0295 |
| $s_{ij}$ (×10$^{-12}$ Pa$^{-1}$) | $s_{33}$=$s_{11}$=8.2, $s_{13}$= -2.6' |
| $g_{ij}$ (×10$^{-10}$C$^{-2}$m$^3$J) | $g_{33}$=5.0 |
| $A^{Si}$ (×10$^{-4}$C$^{-2}$·J) | $A^{S1}$ = 1, $A^{S2}$ = 20000 |
| $F_{ij}$ (×10$^{-11}$C$^{-1}$m$^3$) | $F_{33}$= 3, $F_{13}$= 0 – 3 |
| $W$ (10$^{-30}$m$^3$) | 3 |
| $E_d$ (eV) | −0.1 |
| $N_d^0$ (m$^{-3}$) | 10$^{25}$ |
| $\Sigma$ (eV) | 0.1 |
| Universal constants | $e$=1.6×10$^{-19}$ C, $\varepsilon_0$=8.85×10$^{-12}$ F/m |

Corresponding dependences of the spontaneous polarization, effective piezo-response, average strain and elastic compliance, electron concentration and piezo-conductance on the film thickness *h* are shown in **Figures 2 - 4** Calculated curves appeared very slightly sensitive to the Vegard contribution, which coefficient *W* was varied in the reasonable range (0 – 10) Å$^3$ (compare left (a,c) and right (b,d) columns in **Figures 2-4**). Weak sensitivity to the Vegard strains originated from the donor-blocking boundary conditions used in the 1D numerical modelling, which mean that the full quantity of donors is conserved between the blocking interfaces. The condition minimizes the pure Vegard contribution and does not affect the flexocoupling. Note, that the Vegard contribution to FeMIEC response can be very important in 2D-geometry [12].

Dotted and solid curves, calculated at zero and nonzero flexoelectric coupling constants $F_{ij}$ correspondingly, are very similar at zero external pressure, but become strongly different under external pressure application. At that the difference becomes noticeably stronger for compression ($p_{ext}$>0) than for extension ($p_{ext}$<0). The asymmetry originated from the fact that $p_{ext}$ leads to the linear renormalization of the coefficient $a_{33}$ that becomes $a_{33}^{eff} = \alpha_{33}^T(T-T_c) + 2Q_{33}^{eff} p_{ext}$, where the last term increases or decreases $a_{33}^{eff}$ depending on the $p_{ext}$ sign. The coefficient $a_{33}^{eff}$ defines the critical thickness $h_{cr}$ of the ferroelectricity disappearance in the following way, $h_{cr} = -\frac{g_{33}^{eff}}{a_{33}^{eff}}\left(\frac{1}{\lambda_1 + L_C} + \frac{1}{\lambda_2 + L_C}\right)$ [43], where the renormalized gradient coefficient $g_{33}^{eff} = g_{33} + \frac{2F_{13}^2}{s_{11} + s_{13}}$, correlation and different extrapolation lengths, $L_C = \sqrt{g_{33}^{eff}\varepsilon_0\varepsilon_{33}^b}$ and $\lambda_m = g_{33}^{eff}/A_{33}^{Sm}$ are introduced in **Appendix B** of **Suppl.**

**Mat**. Thus, the flexoelectric coupling renormalizes the gradient coefficient and consequently the extrapolation and correlation lengths [24].

Without flexoelectric coupling all physical quantities depicted in the **Figures 2 - 4** manifest noticeable peculiarities at the critical thickness $h = h_{cr}$. Since $Q_{33}^{eff} > 0$ for PZT, negative $p_{ext}<0$ decreases the critical thickness $h_{cr}$, while positive $p_{ext}>0$ leads to an opposite trend. Therefore $h_{cr}(p_{ext} < 0) < h_{cr}(p_{ext} = 0) < h_{cr}(p_{ext} > 0)$ (compare red, black and blue dashed curves in **Figures 2-4** corresponding to $p_{ext} = -10$ GPa, 0, +10 GPa and $F_{ij} = 0$).

The spontaneous polarization, calculated at $F_{ij} = 0$, emerges at the critical thickness $h_{cr}$, then increases and saturates under the film thickness increasing in a semi-quantitative agreement with the analytical formula $P_3^S = P_S^{bulk}\sqrt{1 - h_{cr}/h}$ [28] (see dashed curves in **Figures 2 a,b**). Effective piezo-response $R_3^{eff}$ calculated at $F_{ij} = 0$ has a divergence at $h = h_{cr}$ and disappears in a paraelectric phase (see dashed curves in **Figures 2 c-d**). The behavior of $R_3^{eff}$ is in agreement with the analytical expression derived in the **Appendix C** of **Suppl.Mat**, $R_3^{piezo} = d_{33}^{PR}\sqrt{1 - \frac{h_{cr}}{h}\left(\frac{\theta(h_{cr}/h)}{|1 - h_{cr}/h|} + \frac{\varepsilon_{33}^b}{\varepsilon_{33}^{sm}}\right)}$, where the piezoresponse amplitude $d_{33}^{PR} \approx 2\varepsilon_0\varepsilon_{33}^{sm}P_S Q_{33}^{eff}$ and the function $\theta(h_{cr}/h) = 2$ at $h < h_{cr}$ and $\theta(h_{cr}/h) = 1$ at $h \geq h_{cr}$.

Apparently nonzero flexoelectric coupling and external pressure induce together polarizations $P_m^{BI} = F_{33}^{eff} p_{ext}/A_{33}^{Sm}$ at the film surfaces (see **Appendix B** of **Suppl.Mat**). The "surface" polarizations cause the built-in field $E^{BI} \sim (P_1^{BI} - P_2^{BI})/h \sim F_{33}^{eff} p_{ext}/h$ that destroys the thickness-induced phase transition to a paraelectric phase at $h = h_{cr}$ and instead induces an electret-like state with irreversible spontaneous polarization at $h < h_{cr}$ (see solid curves in **Figures 2 a-b**). Piezo-response $R_3^{eff}$ calculated from Eq.(6) appeared nonzero in the electret-like state at $h < h_{cr}$ and monotonically decreases with decreasing $h$ (see solid curves in **Figures 2 c-d**).

The spontaneous average strain $\langle u_{33}\rangle$ calculated for $F_{ij} = 0$ and $p_{ext} = 0$ emerges at the critical thickness $h_{cr}$, then it increases and saturates under the film thickness increasing in accordance to the law $\sqrt{1 - h_{cr}/h}$. Nonzero pressure shifts the strain by a constant value $-s_{33}^{eff} p_{ext}$ in accordance with Eq.(4) (compare different dashed curves in **Figure 3 a,b**). Being the derivative of the strain with respect to the applied pressure, effective compliance $S_{33}^{eff}$, calculated at $F_{ij} = 0$ from Eq.(6), has a sharp maximum at $h = h_{cr}$ and drops to a constant value $s_{33}^{eff}$ in the paraelectric phase (see dashed curves in **Figure 3 c-d**).

The built-in field, produced by the joint action of flexocoupling and external pressure, destroys the thickness-induced phase transition at $h = h_{cr}$ and, rather unexpectedly, induces a noticeable increase of the absolute value of strain $|\langle u_{33} \rangle|$ under the film thickness decrease below $h_{cr}$ (see solid curves in **Figure 3 a-b**). It appears that the increase is caused by the flexoelectric term $(P_3(h) - P_3(0))F_{33}^{eff}/h$ in Eq.(5) that scales as $1/h$ at small thicknesses. The term is conditioned by different build-in surface polarizations and can be estimated as $(P_2^{BI} - P_1^{BI})F_{33}^{eff}/h$. Flexoeffect leads to the very pronounced increase of the compliance $S_{33}^{eff}$ with thickness decrease at $h < h_{cr}$ (see solid curves in **Figures 3 c-d**).

Without flexoelectric coupling the average electron concentration $\langle n \rangle$ starts to differ from the equilibrium bulk value $n_0 = N_C F_{1/2}((E_F - E_C)/k_B T)$ for film thickness $h > h_{cr}$, because the spontaneous polarization appears above the critical thickness and start to affect on $\langle n \rangle$ via the deformation potential and depolarization field that is produced by the $div(\vec{P}^S)$. For $p_{ext} = 1$ GPa concentration $\langle n \rangle$ rapidly becomes by one order of magnitude greater than $n_0$ at $h > h_{cr}$ and then saturates under the film thickness increasing. For $p_{ext} = 0$ the concentration $\langle n \rangle$ becomes about one order of magnitude smaller than $n_0$ at $h > h_{cr}$, while it gets two orders of magnitude smaller than $n_0$ at $h > h_{cr}$ for $p_{ext} = -1$GPa, then it reaches a very flat minimum and subsequently slightly increases under the film thickness increasing (see dashed curves in **Figures 4 a,b**). Effective piezo-conductance $\Omega_p$ calculated from Eq.(8) at $F_{ij} = 0$ has a divergence at $h = h_{cr}$ and abruptly disappears in a paraelectric phase at $h < h_{cr}$ (see dashed curves in **Figures 4 c-d**). The pressure induced changes of electron concentration are related with the linear renormalization of the coefficient $a_{33}^{eff}$ by the pressure, $a_{33}^{eff} = \alpha_{33}^T(T - T_c) + 2Q_{33}^{eff} p_{ext}$, since the amplitude of the spontaneous polarization $\vec{P}^S$ depends on $a_{33}^{eff}$ in accordance with LGD-type Euler-Lagrange equation (A.2) listed in the **Appendix B** of **Suppl.Mat.**

The built-in field $E^{BI} \sim F_{33}^{eff} p_{ext}/h$ induces noticeable deviation of $\langle n \rangle$ from the value $n_0$ for all film thicknesses $h$ including the range of small thickness $h \leq h_{cr}$. Furthermore, two peculiarities are present on the thickness dependence of $\langle n \rangle$, namely flat extrema at $h \approx h_{cr}$ followed by inflexion point and then by a sharp drop to $n_0$ value under the film thickness decrease (see solid curves in **Figures 4 a-b**). Therefore effective piezo-conductance $\Omega_p$, being the pressure derivative of $\langle n \rangle$ in accordance with Eq.(8), is nonzero for all film thicknesses $h$ and reveals non-trivial thickness dependence at $p_{ext} \neq 0$ (see solid curves in **Figures 4 c-d**). For $p_{ext} = 1$GPa the piezo-conductance thickness dependence, $\Omega_p(h)$, has two maxima, the first maximum is smeared and located at $h \approx h_{cr}$, another one is flat and located at $h < h_{cr}$. They are separated by a sharp drop (by an order of magnitude), which position corresponds to the inflection point of $\langle n \rangle$. For $p_{ext} = 0$ the dependence $\Omega_p(h)$ has one sharp maximum at $h = h_{cr}$ followed by an inflexion point; after that the rapid decrease of the dependence $\Omega_p(h)$ occurs with $h$ decrease. For $p_{ext} = -1\text{GPa}$ $\Omega_p(h)$ reaches a plateau at $h < h_{cr}$ that continues up to the ultra-small thickness. The physical origin of the non-trivial peculiarities of the effective piezo-conductance thickness dependence is the interplay of the $h$-dependent built-in field and polarization contributions to the electronic state.

Note that the biggest differences $\langle n(p_{ext} > 0) - n(p_{ext} < 0) \rangle$ and $\langle \Omega_p(p_{ext} > 0) - \Omega_p(p_{ext} < 0) \rangle$ (more than 3 orders of magnitude for the pressure difference 2 GPa) correspond to the film thickness $h \sim h_{cr}$ (see vertical green double arrows in **Figures 4a-b**). The changes of $\langle n \rangle$ by orders of magnitude under application of positive and negative pressures can indicate the appearance of high-conductivity (HC) and low-conductivity (LC) states in a thin film with thickness a bit higher than $h_{cr}$, in which swing can be ruled by flexoelectric coupling. Using the analogy with mechanical control of electro-resistive switching in MIECs (piezo-chemical effect) [8], the predicted effect make it possible to control the non-volatile electro-resistive switching in FeMICs by the size effect facilitated by the joint action of external pressure and flexoelectric coupling.

The impact of the Vegard mechanism on the size effects is weak in comparison with the flexoelectric coupling, but the thickness dependence of the piezo-conductance allows one to see the difference between $W = 0$ and $W = 3$ Å$^3$ by comparison of **Figure 4 c** and **4d**.

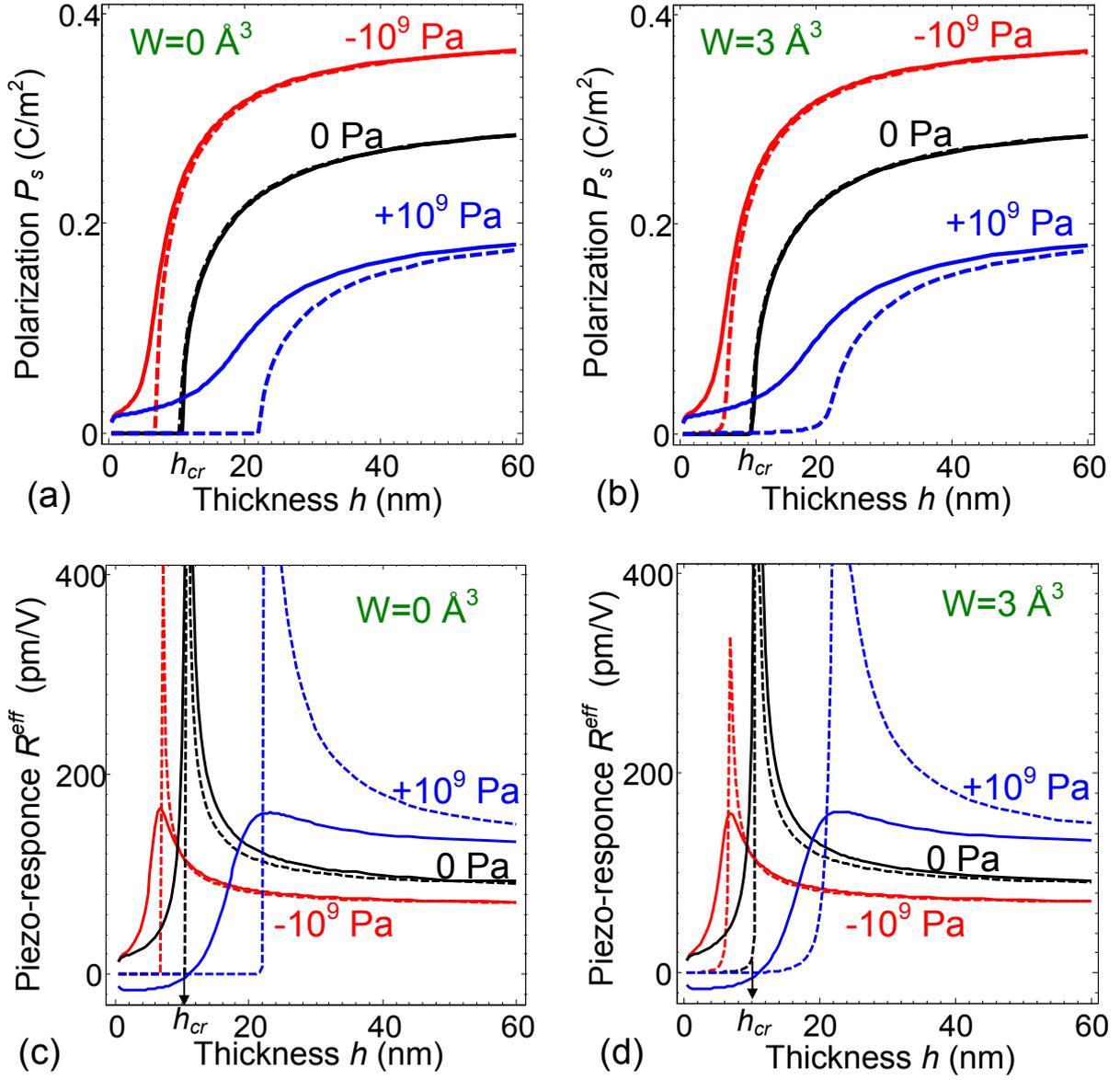

**Figure 2.** Thickness dependence of the average spontaneous polarization $P_3^S$ **(a, b)** and effective piezo-response $R_3^{eff}$ **(c, d)** of ferroelectric PbZr$_{0.5}$Ti$_{0.5}$O$_3$ calculated at RT for different values of external pressure $p_{ext} = -10$ GPa, 0, +10 GPa (shown near the curves) and flexoelectric coefficients $F_{13}=F_{33}=0$ (dashed curves); $F_{13}=1\times10^{-11}$ m$^3$/C, $F_{33}=3\times10^{-11}$ m$^3$/C (solid curves). Vegard coefficient is $W=0$ Å$^3$ **(a, c)** and $W=3$ Å$^3$ **(b, d)**. Other parameters are listed in **Table I**.

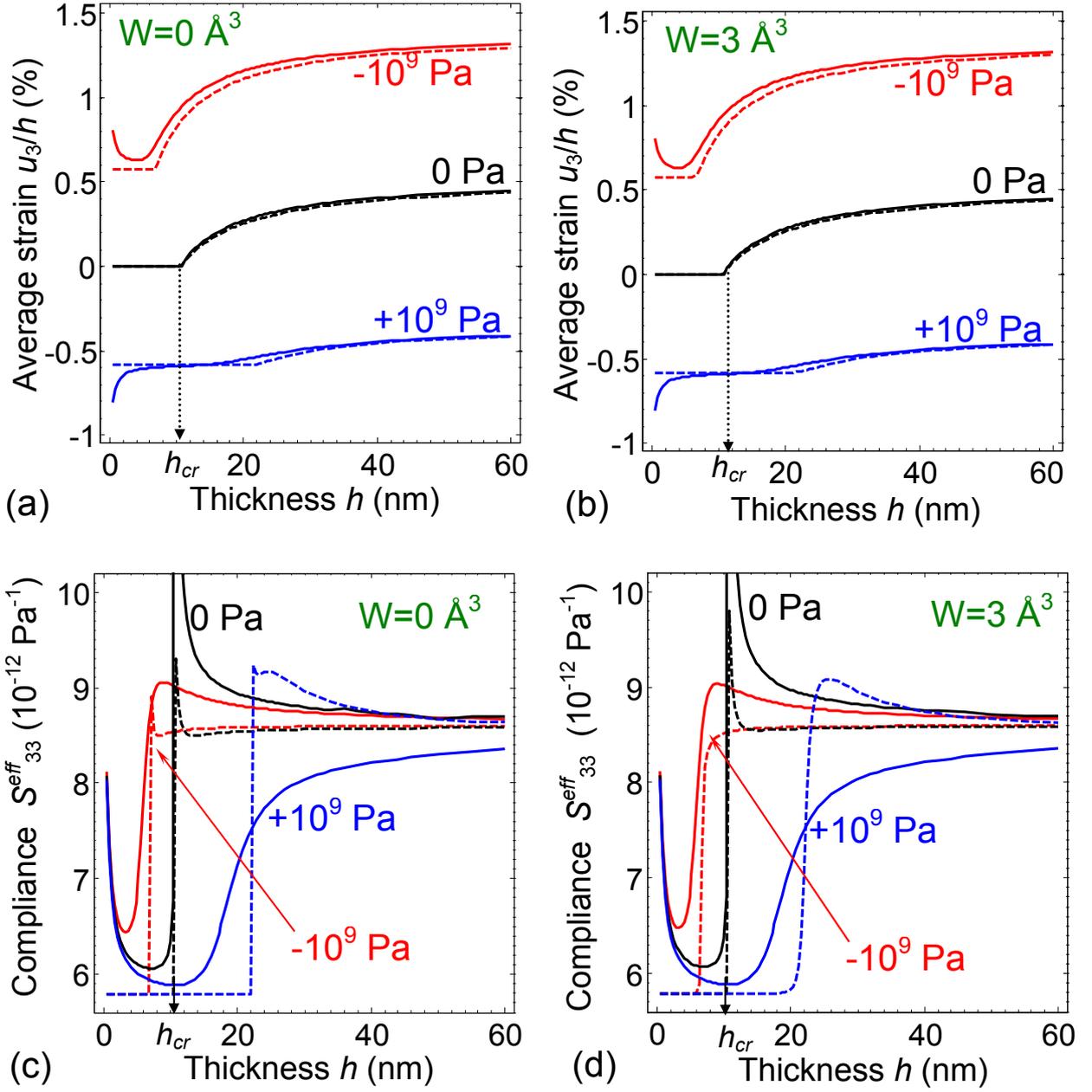

**Figure 3.** Thickness dependence of average strain $\langle u_{33} \rangle$ **(a, b)** and effective elastic compliance $S_{33}^{eff}$ **(c, d)** of ferroelectric PbZr$_{0.5}$Ti$_{0.5}$O$_3$ calculated at RT for different values of external pressure $p_{ext} = -10$ GPa, 0, +10 GPa (shown near the curves) and flexoelectric coefficients $F_{13}=F_{33}=0$ (dashed curves); $F_{13}=1\times10^{-11}$ m$^3$/C, $F_{33}=3\times10^{-11}$ m$^3$/C (solid curves). Vegard coefficient is W=0 Å$^3$ (a, c) and W=3 Å$^3$ (b, d). Other parameters are listed in **Table I**.

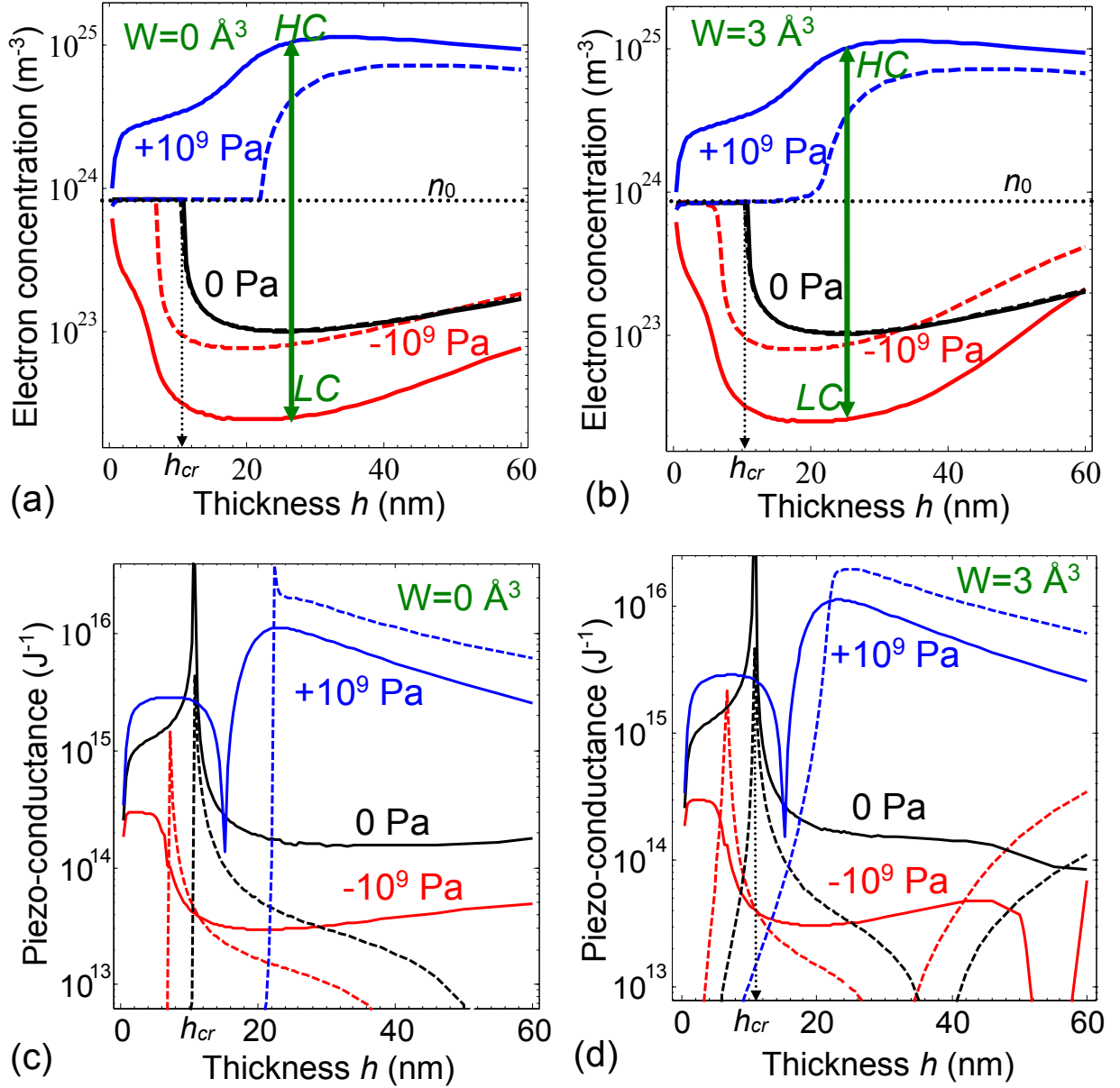

**Figure 4.** Thickness dependence of the average electron concentration $\langle n \rangle$ **(a, b)** and effective piezo-conductance $\Omega_p$ **(c, d)** of ferroelectric PbZr$_{0.5}$Ti$_{0.5}$O$_3$ calculated at RT for different values of external pressure $p_{ext} = -10$ GPa, 0, +10 GPa (shown near the curves) and flexoelectric coefficients $F_{13}=F_{33}=0$ (dashed curves); $F_{13}=1\times10^{-11}$ m$^3$/C, $F_{33}=3\times10^{-11}$ m$^3$/C (solid curves). Green arrows indicate the difference between high-conductivity (HC) and low-conductivity (LC) states. Vegard coefficient is W=0 Å$^3$ (a, c) and W=3 Å$^3$ (b, d). Other parameters are listed in **Table I**.

### IV. Conclusion

Flexocoupling impact on the size effects of the spontaneous polarization, effective piezo-response, elastic strain and compliance, carrier concentration and piezo-conductance have been calculated in thin films of ferroelectric mixed-type semiconductors within LGD-approach combined with classical electrodynamics and semiconductor properties description. Analysis of the self-consistent calculation results revealed that the thickness dependences of aforementioned physical quantities, calculated at zero and nonzero flexoelectric coupling, are very similar without applied pressure, but become strongly different under the application of external pressure $p_{ext}$.

Without flexoelectric coupling the studied physical quantities manifest pronounced peculiarities (disappearance, divergences or sharp maxima, breaks) if the film thickness $h$ approaches the critical thickness $h_{cr}$ of the ferroelectricity existence. We derived analytically how the value of $h_{cr}$ depends on the flexocoupling constants, applied pressure $p_{ext}$, surface energy coefficients and material parameters. Negative pressure $p_{ext}<0$ decreases the critical thickness $h_{cr}$ while a positive one $p_{ext}>0$ leads to an opposite trend.

The combined effect of flexoelectric coupling and external pressure induces the polarizations at the film surfaces. The surface polarizations cause the built-in field that destroys the thickness-induced phase transition to the paraelectric phase at $h = h_{cr}$ and induces the electret-like state with irreversible spontaneous polarization at $h < h_{cr}$. Rather unexpectedly the built-in field leads to the noticeable increase of the average strain and elastic compliance under the film thickness decrease below $h_{cr}$ that scales as $1/h$ at small thicknesses $h$. The increase is conditioned by different build-in surface polarizations at small enough extrapolation lengths, since corresponding built-in field $E^{BI} \sim F_{33}^{eff} p_{ext}/h$ scales as $1/h$ at small thicknesses $h$ ($F_{33}^{eff}$ is the effective flexocoupling constant).

The built-in field induces an essentially non-trivial non-monotonic thickness dependence of free electron density $\langle n \rangle$ for all film thicknesses $h$ including the range of small thickness $h \leq h_{cr}$. Corresponding effective piezo-conductance $\Omega_p$ is nonzero for all film thicknesses $h$ and its thickness dependence is non-monotonic and non-trivial. The physical origin of the peculiarities of the electron concentration and effective piezo-conductance thickness dependences is the interplay of the $h$-dependent built-in field and polarization impact on the electronic state. The impact of the Vegard mechanism on the size effects is weak as anticipated for the donor-blocking boundary conditions, but the thickness dependence of the piezo-conductance is notable.

The changes of $\langle n \rangle$ and $\Omega_p$ by 3 orders of magnitude under application of positive and negative external pressure of 1 GPa can indicate the appearance of high- and low- conductivity states in a thin film with thickness a bit higher than $h_{cr}$, which swing can be ruled by pressure magnitude and flexoelectric coupling. The predicted effect can pave the way for the size effect control of piezo-resistive switching in FeMIECs facilitated by flexoelectric coupling.

Obtained theoretical results can be of fundamental and applied interest for the thin ferroic films physics, semiconductor physics, modern interferometry and Scanning Probe Microscopy development.

Predicted non-trivial behavior of the elastic properties and piezo-conductance are waiting for experimental verification by modern SPM and precise interferometry methods.

**Acknowledgements**

A portion of this research was conducted at the Center for Nanophase Materials Sciences, which is a DOE Office of Science User Facility, CNMS2016-061. E.A.E. and A.N.M. acknowledge National Academy of Sciences of Ukraine (grant 07-06-15). S.V.K. acknowledges Office of Basic Energy Sciences, U.S. Department of Energy. I.S.V. is grateful to the German Research Foundation for support through the grant GE 1171/7-1. M.V.S. acknowledges the grant of the President of the Russian Federation for state support of young Russian scientists-PhD (No. 14.Y30.15.2883-MK) and the project part of the State tasks in the field of scientific activity No. 11.2551.2014/K. Y.K. acknowledges that apportion of this work was supported by Basic Science Research program through the National Research Foundation of Korea funded by the Ministry of Science, ICT & Future Planning (NRF-2014R1A4A1008474).

# Supplementary Materials

## Appendix A. Comment on the form of electron entropy

The electron concentration is

$$n = \int_0^\infty d\varepsilon \cdot g_n(\varepsilon) f\left((\varepsilon - e\varphi - \Sigma_{ij}^e \sigma_{ij} - E_F + E_C)/k_B T\right). \tag{A.1}$$

For the case of parabolic (or effective mass) approximation the density of states $g_n(\varepsilon) \approx \dfrac{\sqrt{2m_n^3 \varepsilon}}{2\pi^2 \hbar^3}$ and thus one get that

$$n = N_C F_{1/2}\left((e\varphi + \Sigma_{ij}^e \sigma_{ij} + E_F - E_C)/k_B T\right), \tag{A.2}$$

where $N_C = \left(m_n k_B T/(2\pi\hbar^2)\right)^{3/2}$ is the effective density of states in the conduction band. Since $E_F = \zeta_e$ in the thermodynamic equilibrium and $\zeta_e = +(\delta G/\delta n) \equiv E_C - \Sigma_{ij}^e \sigma_{ij} - T(\partial S_{el}/\partial n) - e\varphi$, one get that the derivative $\partial S_{el}/\partial n = -k_B F_{1/2}^{-1}(n/N_C)$. Hence the entropy density is $S_{el}[n] = -k_B N_C \int_0^{n/N_C} d\tilde{n} F_{1/2}^{-1}(\tilde{n})$. Here $F_{1/2}^{-1}(\xi)$ is the inverse function to the Fermi ½-integral $F_{1/2}(\xi) = \dfrac{2}{\sqrt{\pi}} \int_0^\infty \dfrac{\sqrt{\zeta}d\zeta}{1+\exp(\zeta-\xi)}$. Note, that in the Boltzmann approximation the electron gas entropy is equal to

$$S_{el}[n] = k_B N_C \left(\frac{n_C}{N_C} - \frac{n_C}{N_C}\ln\left(\frac{n_C}{N_C}\right)\right). \tag{A.3}$$

## Appendix B. Elastic fields and Euler-Lagrange equations

Hereinafter we suppose that the tangential external forces are absent and system is only subjected to external pressure $p_{ext}$ along $x_3$-axis. For the case of mechanical equilibrium equations with boundary conditions are $\partial \sigma_{i3}/\partial x_3 = 0$, $\sigma_{i3}|_S = 0$ ($i$=1, 2) and $\sigma_{33}(h)|_S = -p_{ext}$. Elastic displacement $u_i$ is zero at the surface $z = h$, i.e. $u_i(h) = 0$.

For a ferroelectric film with aforementioned boundary conditions, nonzero elastic field components are the following

$$\sigma_{33} = -p_{ext}, \tag{B.1a}$$

$$\sigma_{11} = \sigma_{22} = \frac{s_{13} p_{ext} - W_{11}^d \delta N_d^+ - \Sigma_{11}^e \delta n}{s_{11} + s_{13}} - \frac{Q_{13} P_3^2}{s_{11} + s_{13}} - \frac{F_{13}}{s_{11} + s_{13}} \frac{\partial P_3}{\partial x_3}, \tag{B.1b}$$

$$u_{33} = s_{33}^{eff} p_{ext} + W_{33}^{eff} \delta N_d^+ + \Sigma_{33}^{eff} \delta n + F_{33}^{eff} \frac{\partial P_3}{\partial x_3} + Q_{33}^{eff} P_3^2. \tag{B.1c}$$

Here we used that $\sigma_{11} = \sigma_{22}$ from the symmetry consideration and introduced the apparent coefficients

$$s_{33}^{eff} = s_{33} - \frac{2s_{13}^2}{s_{11}+s_{13}}, \quad W_{33}^{eff} = W_{33}^d - \frac{2s_{13}W_{11}^d}{s_{11}+s_{13}}, \quad \Sigma_{33}^{eff} = \Sigma_{33}^e - \frac{2s_{13}\Sigma_{11}^e}{s_{11}+s_{13}}, \quad F_{33}^{eff} = F_{33} - \frac{2s_{13}F_{13}}{s_{11}+s_{13}}$$

and $Q_{33}^{eff} = Q_{33} - \frac{2s_{13}Q_{13}}{s_{11}+s_{13}}$. Voigt notations are introduced for the electrostriction $Q_{ij}$, gradient coefficients $g_{ij}$, flexoelectric $F_{ij}$ and elastic compliances $s_{ij}$ tensors, while full matrix notations are used for all other tensors. The tensors components with subscripts 12, 13 and 23 are equal for materials with cubic parent phase. Note, that the piezoelectric contribution is automatically included in the equations (B.1) as linearized electrostriction, since $P_3 \approx P_3^S + \varepsilon_0(\varepsilon_{33}^f - 1)E_3$.

Using Equations (B.1), we excluded the stresses from the equation for polarization $P_3$ and concentration of donors $N_d^+$. Remained equations can be solved numerically. Note, that the substitution of expressions (B.1) in the functional (1a) leads to the appearance of the flexoelectric coupling with Vegard and piezoelectric terms proportional to the products $W_{ii}^{eff} F_{33}^{eff}$, $\Sigma_{ii}^{eff} F_{33}^{eff}$, etc.

Euler-Lagrange equation for determination of the ferroelectric polarization component has an explicit form:

$$\left(a_{33}^{eff} + \frac{4Q_{13}W_{11}^d}{s_{11}+s_{13}}\delta N_d^+\right)P_3 + b_{33}^{eff} P_3^3 - g_{33}^{eff}\frac{\partial^2 P_3}{\partial x_3^2} - \frac{2F_{13}W_{11}^d}{s_{11}+s_{13}}\frac{\partial N_d^+}{\partial x_3} = -\frac{\partial \varphi}{\partial x_3} \quad (B.2)$$

The effective coefficients $g_{33}^{eff} = g_{33} + \frac{2F_{13}^2}{s_{11}+s_{13}}$, $a_{33}^{eff} = \alpha_{33}^T(T - T_c) + 2Q_{33}^{eff} p_{ext}$ and $b_{33}^{eff} = \left(b_{33} + \frac{4Q_{13}^2}{s_{11}+s_{13}}\right)$ are introduced. Boundary conditions for the out-of-plane polarization component $P_3$ are of the third kind [44]:

$$\left(P_3 - \lambda_1\frac{\partial P_3}{\partial x_3} - \frac{2F_{13}W_{11}^d}{s_{11}+s_{12}}\frac{\delta N_d^+}{A_{33}^{S1}} - \frac{2F_{13}Q_{13}}{s_{11}+s_{13}}\frac{P_3^2}{A_{33}^{S1}} - P_1^{BI}\right)\Bigg|_{x_3=0} = 0, \quad (B.3a)$$

$$\left(P_3 + \lambda_2\frac{\partial P_3}{\partial x_3} + \frac{2F_{13}W_{11}^d}{s_{11}+s_{13}}\frac{\delta N_d^+}{A_{33}^{S2}} + \frac{2F_{13}Q_{13}}{s_{11}+s_{13}}\frac{P_3^2}{A_{33}^{S2}} + P_2^{BI}\right)\Bigg|_{x_3=h} = 0. \quad (B.3b)$$

The extrapolation length $\lambda_m = \frac{g_{33}^{eff}}{A_{33}^{Sm}}$ is determined by the surface energy and the surface state. Physically realistic range for $\lambda_m$ is 0.5 – 2 nm [45]. Built-in polarizations at the surfaces $x_3$=0 and $x_3$=h are, respectively, $P_1^{BI} = \frac{F_{33}^{eff} p_{ext}}{A_{33}^{S1}}$ and $P_2^{BI} = \frac{F_{33}^{eff} p_{ext}}{A_{33}^{S2}}$.

**Appendix C. Analytical estimates of effective electromechanical response**

Below we will demonstrate that the flexocoupling can impact strongly on size effect of effective piezo-response in a ferroelectric layer with a strong gradient of mobile defects.

Using the linear approximation for the dependence of polarization on electric field in a tetragonal phase of ferroelectric, $P_k \approx P_k^S + \varepsilon_0(\varepsilon_{kl}^f - 1)E_l$, the piezoelectric strain $u_{ij}^{piezo}$ is proportional to the convolution of the piezoelectric coefficients with electric field, $u_{ij}^{piezo} \approx d_{ijk}^{eff} E_k$ in accordance with Eq.(3), where the apparent piezoelectric coefficient $d_{ijk}^{eff} = 2\varepsilon_0(\varepsilon_{kn}^f - \delta_{kn})Q_{ijnl}P_l^S$ is introduced. Consequently, vertical effective piezoresponse, defined as the derivative of the surface displacement $u_3^{piezo} \approx \int_0^h u_{33}^{piezo} dx_3$ with respect to the applied voltage $V$, is exactly equal to $d_{33}^{eff}$, because $\int_0^h E_3 dx_3 = V$.

When ferroelectric ceramic layer thickness $h$ decreases approaching the critical thickness $h_{cr}$, the apparent piezoelectric coefficient $d_{33}^{eff}$ becomes thickness-dependent, because the ferroelectric polarization decreases and dielectric permittivity component changes with thickness. Without external pressure the surface polarization are zero ($F_{33}^{eff} p_{ext}/A_{33}^{Si} = 0$) and the following expressions are valid:

$$P_3^S = P_S^{bulk}\sqrt{1 - \frac{h_{cr}}{h}}, \qquad \varepsilon_{33}^f(h) = \varepsilon_{33}^b + \frac{\varepsilon_{33}^{sm}\theta(h_{cr}/h)}{|1 - h_{cr}/h|}, \qquad \text{(C.1)}$$

where the function $\theta(h_{cr}/h) = 2$ at $h < h_{cr}$ and $\theta(h_{cr}/h) = 1$ at $h \geq h_{cr}$ in accordance with Curie-Weiss law. According to the expressions (C.1), $P_3^S$ disappears at $h \leq h_{cr}$ and dielectric permittivity component diverges at $h = h_{cr}$. $\varepsilon_{33}^{sm}$ is a soft-mode related relative dielectric permittivity of a bulk ferroelectric material, $\varepsilon_{33}^b$ is a background contributions (typically $\varepsilon_{33}^b \ll \varepsilon_{33}^{sm}$). The critical thickness depends on the ferroelectric material parameters, temperature, extrapolation length $\lambda_P$ and flexoelectric coupling coefficients [24, 28**Ошибка! Закладка не определена.**]:

$$h_{cr} = -\frac{g_{33}^{eff}}{a_{33}^{eff}}\left(\frac{1}{\lambda_1 + L_C} + \frac{1}{\lambda_2 + L_C}\right), \qquad \text{(C.2)}$$

where the correlation length $L_C = \sqrt{g_{33}^{eff}\varepsilon_0\varepsilon_{33}^b}$ is introduced. Extrapolation length $\lambda_m = g_{33}^{eff}/A_{33}^{Sm}$ is positive or zero. Note that $L_C$ can be much smaller than the lattice constant due to depolarization field influence, so that $\lambda_m + L_C \approx \lambda_m$ for extrapolation lengths more than 1 nm. In this case $h_{cr} \approx -\frac{A_{33}^{S1} + A_{33}^{S2}}{a_{33}^{eff}}$ is dependent on external pressure, but independent on the flexocoupling strength.

Allowing for Eqs.(C.1)-(C.2) the thickness dependence of effective piezo-response in a ferroelectric phase has the form:

$$R_3^{piezo} = \frac{\partial u_3^{piezo}}{\partial V} = d_{33}^{PR}\sqrt{1-\frac{h_{cr}}{h}}\left(\frac{\theta(h_{cr}/h)}{|1-h_{cr}/h|}+\frac{\varepsilon_{33}^b}{\varepsilon_{33}^{sm}}\right). \qquad (C.3)$$

In accordance with Eq.(B.1c) the piezoresponse amplitude $d_{33}^{PR}$ is a combination of tabulated piezoelectric coefficients of a bulk ferroelectric material, $d_{33}^{PR} \approx 2\varepsilon_0\varepsilon_{33}^{sm}P_S Q_{33}^{eff}$. The piezoresonse $R_3^{piezo}$ disappears in a paraelectric phase, at $h \leq h_{cr}$, because of $P_S$ disappearance accordingly to Eq.(C.1).

Note that the expression (C.3) is valid at zero flexoelectric coupling or/and external pressure, because for this case surface polarizations in Eqs.(B.3) are absent.

In accordance with Eq.(3) the Vegard strain $u_{ij}^{chemo}$ is proportional to the convolution of the apparent Vegard coefficients $W_{ij}^{eff}$ with mobile defect concentration spatial variation $\delta N_d^+(\mathbf{x})$, $u_{ij}^{chemo} \approx W_{ij}^{eff}\delta N_d^+(\mathbf{x})$. Deformation potential leads to the strain $u_{ij}^{def} \approx \Sigma_{ij}^{eff}\delta n(\mathbf{x})$. The flexoelectric strain $u_{ij}^{flexo}$ is proportional to the convolution of the apparent flexoelectric coefficients $F_{ijkl}^{eff}$ with polarization gradient, $u_{ij}^{flexo} \approx F_{ijkl}^{eff}\partial P_k/\partial x_l$.

In a paraelectric phase $u_{33}^{flexo} \approx F_{33}^{eff}\varepsilon_0(\varepsilon_{33}^f-1)\partial E_3/\partial x_3$, because $P_3 \approx \varepsilon_0(\varepsilon_{3n}^f-\delta_{3n})E_n$ at $h \leq h_{cr}$ and small applied voltages. Approximate analytical expressions for the space charges, electric potential and field can be derived within the linear Debye approximation valid at very small applied voltages, $|eZ_d\varphi/k_B T| \ll 1$, namely

$$\delta N_d^+ \approx n_0\left(\exp\left(\frac{eZ_d\varphi}{k_B T}\right)-1\right) \approx \frac{\varepsilon_0\varepsilon_{33}^f\varphi}{eh_d^2}, \quad \delta n \approx n_0\left(\exp\left(\frac{-e\varphi}{k_B T}\right)-1\right) \approx -\frac{\varepsilon_0\varepsilon_{33}^f\varphi}{eh_d^2} \qquad (C.4a)$$

$$\rho \approx \frac{\varepsilon_0\varepsilon_{33}^f\varphi}{h_d^2}, \quad \varphi = -V\frac{\sinh((x_3-h)/h_d)}{\sinh(h/h_d)}, \quad E_3 = \frac{V}{h_d}\frac{\cosh((x_3-h)/h_d)}{\sinh(h/h_d)} \qquad (C.4b)$$

The introduced screening length $h_d = \sqrt{k_B T/(\varepsilon_0\varepsilon_{33}^f e^2(Z_d+1)n_0)}$ is thickness-dependent, because $\varepsilon_{33}^f$ is thickness-dependent according to Eq.(C.1).

Using Eqs.(C.4) the flexoelectric and chemical contributions to the longitudinal effective electromechanical response $R_3^{flexo+chemo}$ can be estimated in the paraelectric phase as:

$$R_3^{flexo+chemo} = \frac{\partial}{\partial V}\left(\int_0^h (u_{33}^{flexo}+u_{33}^{chemo}+u_{33}^{def})dx_3\right) \approx R_{33}^{FC}\frac{\varepsilon_{33}^f(h)}{\varepsilon_{33}^{sm}}\frac{1-\cosh(h/h_d)}{h_d\sinh(h/h_d)} \qquad (C.5)$$

In accordance with Eq.(4c) the response amplitude $R_{33}^{FC}$ is a sum of three contributions, flexoelectric, Vegard and deformation potential tensors, $R_{33}^{FC} \approx \varepsilon_0 \varepsilon_{33}^{sm} \left( F_{33}^{eff} - \frac{W_{33}^{eff} - \Sigma_{33}^{eff}}{e} \right)$.

In order to calculate the quadratic electrostrictive contribution that can dominate in paraelectric phase, we used the same assumptions as for derivation of Eq.(C.5). The electrostrictive displacement in the paraelectric phase is $u_3^Q(h) = Q_{33}\left(\varepsilon_0\left(\varepsilon_{33}^f - 1\right)\right)^2 \int_0^h E_3^2 dx_3$. The displacement $u_3^Q$ is proportional to $V^2$, because $E_3^2 \sim V^2$ in accordance with Eq.(C.4b). Hence $u_3^Q$ becomes negligibly small in comparison with the flexoelectric and chemical ones electrostrictive contribution at small enough $V$.

### Appendix D. Analytical estimates of effective piezo-conductance

In order to estimate analytically the effective piezo-conductance, defined as the dependence of electro-conductance on applied pressure, one should perform rigorous calculations of the electric current and its derivatives on applied voltage and pressure. In the considered 1D case of donor-blocking and electron-open electrodes the donor current is zero and the electron current is constant. Namely:

$$-eZ_d \eta_d N_d^+ \frac{\partial \zeta_d}{\partial x_3} = 0, \quad e\eta_e n \frac{\partial \zeta_e}{\partial x_3} = J. \tag{D.1}$$

Let us try to solve Eqs.(D.1) in the Boltzmann-Plank-Nernst (BPN) approximation for donors' and electrons' concentrations. The approximation for donors means that $\ln\left(N_d^+ / \left(N_d^0 - N_d^+\right)\right) \approx \ln\left(N_d^+ / N_d^0\right)$, $N_d^+ \approx N_d^0 \exp\left(\left(-eZ_d\varphi + E_d + W_{ij}^d \sigma_{ij} - \zeta_d\right)/k_B T\right)$ and $\zeta_d = const$. BPN approximation for electrons means that $n \approx N_C \exp\left(\left(e\varphi + \Sigma_{ij}^e \sigma_{ij} - E_C + \zeta_e\right)/k_B T\right)$ and $\zeta_e = E_C - \Sigma_{ij}^e \sigma_{ij} - e\varphi + k_B T \ln(n/N_C)$. One can solve the equations for electron concentration $e\eta_e n \left( \frac{k_B T}{n} \frac{\partial n}{\partial x_3} - e\frac{\partial \varphi}{\partial x_3} - \Sigma_{ij}^e \frac{\partial \sigma_{ij}}{\partial x_3} \right) = J$ in the approximation $\frac{\partial n}{\partial x_3} = \frac{J}{e\eta_e k_B T} + \frac{\Sigma_{ij}^e}{k_B T} \bar{n} \frac{\partial \sigma_{ij}}{\partial x_3} + \frac{e\bar{n}}{k_B T} \frac{\partial \varphi}{\partial x_3}$ and obtain the expression $n = \frac{J x_3}{e k_B T \eta_e} + \frac{\bar{n}}{k_B T}\left(e\varphi + \Sigma_{ij}^e \sigma_{ij}\right) + C$. From here $C = \bar{n}\left(1 - \frac{1}{k_B T}\left(e\bar{\varphi} + \Sigma_{ij}^e \bar{\sigma}_{ij}\right)\right) - \frac{Jh}{2ek_B T \eta_e}$ and so

$$n = \frac{J(x_3 - h/2)}{ek_B T \eta_e} + \bar{n}\left(1 + \frac{e(\varphi - \bar{\varphi}) + \Sigma_{ij}^e\left(\sigma_{ij} - \bar{\sigma}_{ij}\right)}{k_B T}\right). \tag{D.2}$$

From the boundary conditions $n(0) = n_0$ and $n(h) = n_1$ one can determine the values of $J$ and $\bar{n}$, namely the solution is

$$\bar{n} = (n_1 + n_0)\left(2\left(1 - \frac{\bar{\varphi} + \Sigma_{ij}^e \bar{\sigma}_{ij}}{k_B T}\right) + \frac{e(\varphi(h) + \varphi(0)) + \Sigma_{ij}^e(\sigma_{ij}(h) + \sigma_{ij}(0))}{k_B T}\right)^{-1} \quad \text{(D.3a)}$$

$$J = \frac{ek_B T \eta_e}{h}\left(n_1 - n_0 - \bar{n}\left(\frac{e(\varphi(h) - \varphi(0)) + \Sigma_{ij}^e(\sigma_{ij}(h) - \sigma_{ij}(0))}{k_B T}\right)\right) \quad \text{(D.3b)}$$

Since $\varphi(0) - \varphi(h) = V$ and we can regard that $n_1 = n_0$, so the current $J = e\eta_e \bar{n}\left(\frac{eV + \Sigma_{ij}^e(\sigma_{ij}(0) - \sigma_{ij}(h))}{h}\right)$ and $\bar{n} = 2n_0\left(2\left(1 - \frac{\bar{\varphi} + \Sigma_{ij}^e \bar{\sigma}_{ij}}{k_B T}\right) + \frac{eV + \Sigma_{ij}^e(\sigma_{ij}(h) + \sigma_{ij}(0))}{k_B T}\right)^{-1}$. So that the piezo-conductance defined as the derivative of the conductance $\Omega_p$ on applied pressure, $\frac{d\Omega_p}{dp_{ext}} \equiv \frac{1}{E_{ext}} \frac{dJ}{dp_{ext}}$, where we regard that $E_{ext} = V/h$, acquires the form:

$$\frac{d\Omega_p}{dp_{ext}} \approx e\eta_e \frac{d}{dp_{ext}}\left(e\bar{n} + \frac{\Sigma_{ij}^e(\sigma_{ij}(0) - \sigma_{ij}(h))}{V}\bar{n}\right) \quad \text{(D.4a)}$$

Here the convolution $\Sigma_{ij}^e \sigma_{ij} = -\left(\Sigma_{33}^e - \frac{2s_{13}\Sigma_{11}^e}{s_{11} + s_{13}}\right)p_{ext} - \frac{2\Sigma_{11}^e}{s_{11} + s_{13}}\left(Q_{13}P_3^2 + F_{13}\frac{\partial P_3}{\partial x_3}\right)$ in accordance with Eqs.(B.1), and for small deformation potentials or slowly-varying stress field the second term in brackets can be neglected, resulting in

$$\frac{d\Omega_p}{dp_{ext}} \approx e^2 \eta_e \frac{d\bar{n}}{dp_{ext}}. \quad \text{(D.4b)}$$

Let us calculate the concentration derivatives on the applied pressure. When the system is in thermodynamic equilibrium, currents are absent and electrochemical potentials are equal to Fermi level. For the case donor concentration is $N_d^+ = N_d^0\left(1 - f\left((E_d + W_{ij}^d \sigma_{ij} - eZ_d\varphi + E_F)/k_B T\right)\right)$, where $E_F$ is the Fermi energy level in equilibrium. Electron density is $n = N_C F_{1/2}\left((e\varphi + \Sigma_{ij}^e \sigma_{ij} + E_F - E_C)/k_B T\right)$. Hence the concentration derivatives on the applied pressure are:

$$\frac{\partial n}{\partial p_{ext}} = N_C F_{1/2}'\left(\frac{Y}{k_B T}\right)\frac{1}{k_B T}\frac{\partial Y}{\partial p_{ext}} \quad \text{(D.5a)}$$

Where the function $F_{1/2}'(\xi) = \frac{2}{\sqrt{\pi}} \int_0^\infty \frac{\sqrt{\zeta}\exp(\zeta - \xi)d\zeta}{(1 + \exp(\zeta - \xi))^2}$. In accordance with expressions (B.1), the arguments

$$Y = -\Sigma_{33}^{eff} p_{ext} - \frac{2\Sigma_{11}^e}{s_{11} + s_{13}}\left(Q_{13}P_3^2 + F_{13}\frac{\partial P_3}{\partial x_3}\right) + e\varphi + E_F - E_C \quad \text{(D.6a)}$$

[33] Note that we did not include the higher elastic gradient term, $\frac{1}{2}v_{ijklmn}(\partial\sigma_{ij}/\partial x_m)(\partial\sigma_{kl}/\partial x_n)$, in the functional (1a), because its value and properties are still under debate. Therefore we are subjected to use only one half ($F_{ijkl}P_k(\partial\sigma_{ij}/\partial x_l)$) of the full Lifshitz invariant $F_{ijkl}(P_k(\partial\sigma_{ij}/\partial x_l)-\sigma_{ij}(\partial P_k/\partial x_l))/2$. The higher elastic gradient term is required for the stability of the functional with full Lifshitz invariant included. The usage of either the term $F_{ijkl}P_k(\partial\sigma_{ij}/\partial x_l)$ or the term $F_{ijkl}(P_k(\partial\sigma_{ij}/\partial x_l)-\sigma_{ij}(\partial P_k/\partial x_l))/2$ does not affect on the equations of state, but influences on the elastic boundary conditions.

[38] Note, that the total polarization component $P_k^t(\mathbf{r})$ includes both ferroelectric contribution $P_k(\mathbf{r})$, originated from a soft mode, and non-ferroelectric one, $P_k^b(\mathbf{r})$. For both ferroelectrics and highly-polarized paraelectrics the non-ferroelectric contribution is typically much smaller than the ferroelectric one, $|P_k(\mathbf{r})| \gg |P_k^b(\mathbf{r})|$. So that we can regard that $P_k^t(\mathbf{r}) \approx P_k(\mathbf{r})$ and hereinafter omit the subscript "t" in the term $Q_{ijkl}P_k P_l$ in Eq.(1a).

[43] The approximate expression is valid with high accuracy at small concentration of free carriers